# Automated Attack and Defense Framework for 5G Security on Physical and Logical Layers

Zhihong Tian[1], Yanbin Sun[1]*, Shen Su[1], Mohan Li[1], Xiaojiang Du[2], *Senior Member, IEEE*, and Mohsen Guizani[3], *Fellow, IEEE*

**Abstract**

The 5th generation (5G) network adopts a great number of revolutionary technologies to fulfill continuously increasing requirements of a variety of applications, including ultra-high bandwidth, ultra-low latency, ultra-massive device access, ultra-reliability, and so on. Correspondingly, traditional security focuses on the core network, and the logical (non-physical) layer is no longer suitable for the 5G network. 5G security presents a tendency to extend from the network center to the network edge and from the logical layer to the physical layer. The physical layer security is also an essential part of 5G security. However, the security of each layer in 5G is mostly studied separately, which causes a lack of comprehensive analysis for security issues across layers. Meanwhile, potential security threats are lack of automated solutions. This article explores the 5G security by combining the physical layer and the logical layer from the perspective of automated attack and defense, and dedicate to provide automated solution framework for 5G security.

## 1 Introduction

THE wireless network has become an increasingly indispensable part of current and future life. With the rapid development of information technologies such as the mobile Internet, Internet of Things (IoTs), Internet of Vehicles (IoVs), etc., a great number of new applications have emerged and are widely applied through wireless communications. By 2022, global mobile devices will grow from 8.6 billion in 2017 to 12.3 billion by 2022, and the traffic from wireless and mobile devices will be about 77.5 exabytes per month [1]. Although the 4G network and previous wireless networks can barely satisfy the data traffic and device access requirements for the next few years, the future requirements cannot be satisfied, especially for large IoT devices, high mobile devices, and high-traffic applications. To support mobile wireless communication in the future, the 5G network is emerged and serves as a key enabler to support the next generation wireless communication.

The 5G network aims to achieve "internet of everything", and it adopts a significant number of revolutionary technologies to support diversified applications and services, such as VR, AR, IoT, IoV, the device to device communications, e-healthcare, machine to machine communications and Financial Technology (FinTech) [2]. On the physical layer, the full-duplex (FD)

1 Cyberspace Institute of Advanced Technology, Guangzhou University, Guangzhou 510006, China
2 Department of Computer and Information Sciences, Temple University, Philadelphia, USA
3 College of Engineering, Qatar University, Qatar
* Corresponding author: Yanbin Sun (sunyanbin@gzhu.edu.cn)

communication, massive multiple-input-multiple-output (MIMO), ultra-dense network (UDN) and millimeter wave (mmWave) are adapted to acquire high data rate, low latency, massive connectivity, and high coverage rate. On the logical layer, network functions virtualization (NFV), software-defined network (SDN), cloud-based network (C-RAN), routing model [3] are adapted to support network virtualization and network slicing, which is helpful to provide flexible, efficient and low-cost services. Based on above technologies, the future 5G network is designed to support three main scenarios: enhanced mobile broadband (eMBB), massive machine-type communication (mMTC) and ultra-reliable and low latency communication (uRLLC) [4].

New requirements, technologies, and scenarios bring new challenges to 5G security and demand complex and diverse security mechanisms. The research of 5G security should be treated as part of 5G architecture rather than the "patch" of 5G architecture, such that the security design is integrated into the 5G architecture. On the one hand, the security of 5G is guaranteed, and the number of network attack is greatly reduced. On the other hand, the patching solution is avoided to reduce the bloat of 5G architecture.

5G security has drawn lots of attentions. Key security problems of 5G are also widely studied. Existing 5G security researches can be divided into two aspects: the physical layer security and the logical layer security. The physical layer security provides encryption, authentication, cryptographic key distribution and management via various technologies of physical layer [5]. The logic layer security focuses on the security of NFV and SDN, e.g., security isolation, security of inter-slice communications, configuration errors, virtualization threats, hypervisor hijacking, and so on [6].

However, there are still some issues for 5G security. (1) Physical layer security and logical layer security are loosely coupled, which causes the cross-layer security issue to be ignored or insufficiently studied. Since the new application of 5G network relies on both the physical layer and the logical layer, a security issue may be caused by vulnerabilities on multiple layers rather than a single layer. (2) Besides existing security issues, potential security threats, such as the 0-day vulnerability, unknown attacks, should also be concerned. How to find these vulnerabilities and attacks plays an important role in 5G security. (3) The automated threat detection and response mechanism is lacking in 5G security. Most existing security solutions depend on the manual intervention which is difficult to be applied to the large-scale and complex 5G networks due to the efficiency and accuracy requirements.

Therefore, it is crucial to highlight the automated solutions for security issues of 5G networks. This article is dedicated to finding a solution for 5G security based on automated attack and defense and integrate the automated security mechanism into 5G architecture. The rest of the article is organized as follows. We first discuss the security challenges of 5G on a hierarchical framework combining the physical layer and the logical layer, and then propose a hierarchical attack and defense model of 5G. Based on the model, a security framework for 5G is proposed from the perspective of automated attack and defense.

## 2 Security Challenges on Hierarchical 5G Framework

The security challenge of the 5G network has been reviewed by many survey and tutorial articles [5, 6, 7]. Most of these articles focus on the security challenges of 5G new technologies. In order to support the automated attack and defense framework, the security challenges of 5G are analyzed in the perspective of attack/defense object. We first summarize a universal hierarchical framework for 5G networks, and then the security challenge is analyzed.

**2.1 Hierarchical Framework of 5G Networks**

Although the standard 5G architecture is not unified, a universal 5G framework can still be summarized according to existing 5G architectures. As shown in Figure 1, the 5G network is divided into two types of layers: the physical layer and the logical layer. The right part containing SDN, network slicing, and corresponding management and orchestration can be integrated into each layer. The physical layer adopts physical devices and new communication technologies (massive MIMO, FD, mmWave, etc.) to construct wireless communication network for upper (logical) layers. The logical layer uses the physical resource provided by the underlying network via some networking technologies (network slicing, SDN, NFV, clouding, etc.) to support new applications.

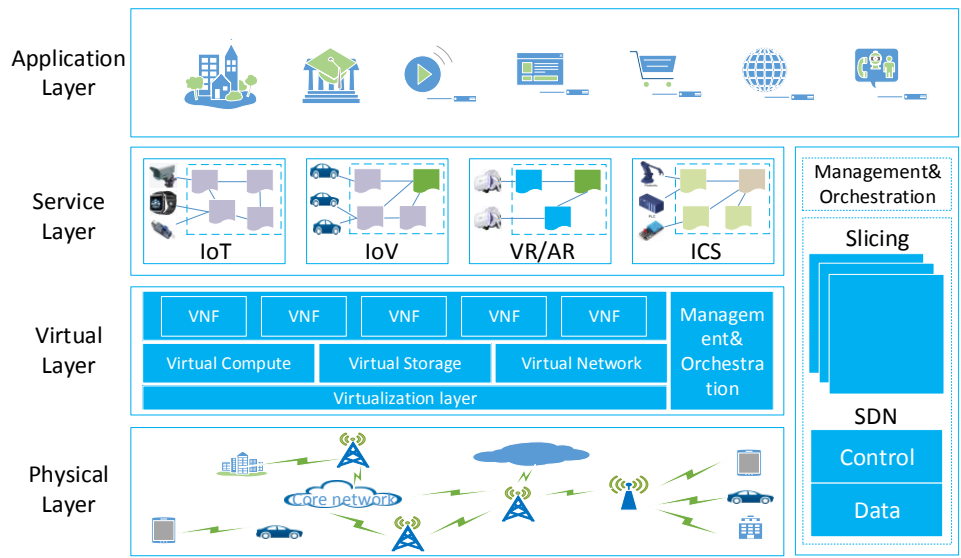

Figure 1. Hierarchical Framework of 5G Network

According to the layer function, the logical layer can continue to be divided into multiple layers including the virtual layer, service (slice) layer and application layer. The virtual layer adopts the virtualization technology to transform the different physical devices to unified virtual resources and transfer the network function from hardware-based applications to software-based applications [8]. The service layer obtains the resource (storage, computing or communication) from the virtual layer via network slicing, such that the resource of service for a specific application scenario can be satisfied. The application layer supports the specific task based on the resource provided by the service layer, e.g., the smart city is one of the specific applications of IoT service.

**2.2 Security Challenges and Solutions**

Instead of reviewing the security challenges of technologies, we focus on the security challenges of attack/defense object. By analyzing and combining the attack/defense objects of each layer, the objects are shown in Table 1. Note that, the object in the table is only a rough division. Some objects can continue to be divided, e.g., the physical device can be divided into multiple hardware elements, such as CPU, GPU, memory, and so on. It is not further discussed here.

Table 1. Objects of 5G networks

| Category | Object |
|---|---|

| Hardware device | Base station (BS), Repeater, Router, Switch, Server, Smart device, etc |
|---|---|
| Channel | Wireless channel, Optical fiber, Cable, etc. |
| Virtual entity | Openstack, VMWare, Hypervisor, Docker, vSwitch, Virtual machine (VM), Virtual resource, etc. |
| Operating system (OS) | Enea OSE, μC/OS, Linux, Windows, etc |
| Control software | SDN controller, Slicing controller, etc. |
| Application software | Web browser, Database, Smartphone app, etc. |
| Protocol | Communication protocol, Networking protocol, Authorization protocol, API, etc. |

The physical layer security focuses on the hardware device and the wireless channel. The main challenges of hardware device are malicious destruction, side channel attacks, and fake devices. The first two issues cause unavailability and data leakage for the device and can be solved by the security management to prevent attackers from getting close to the device. The fake device (fake BS and malicious repeater) causes the man-in-the-middle attack and can be solved by authorization and encryption mechanisms. For the wireless channel, the eavesdropping attack and unauthorized access, which are caused by the open nature of radio propagation [9], are two main challenges. Encryption and authorization mechanisms can also be used to solve the two challenges. The inherent security attributes of the wireless channel, including unique, diverse and reciprocity, make the wireless channel cannot be measured, reconstructed, and replicated by the third party. Based on these attributes, enhanced encryption and authentication mechanisms are obtained by using wiretap coding technologies, secure multi-antenna technologies, CSI/RSS/phase/code-based key generation technologies, wiretap code-based authentication, RF recognition approaches, and so on [10].

On the logical layer, the physical resources are virtualized and sliced to support services. There are three main security challenges. (1) Data leakage caused by the shared resource. Although the resources of different objects are logically separated, they may belong to the same physical device. The re-allocation of resources may transfer the resource from a user to another user and lead to data leakage. Meanwhile, poor isolation strategies may also lead to data leakage or attack. Approaches such as VM/slice isolation, authorization, data erasure, encryption and secured communication protocol (TLS, SSH) [9] are promising solutions. (2) Vulnerabilities of software and protocol. The main focus of the 5G network is the management and orchestration software, i.e., controller. The design of the controller is often centralized. Thus, an attacker can acquire high permissions by the exploits of controller vulnerabilities, and then launches attacks at the corresponding layer. Besides the controller vulnerability, vulnerabilities of OS, application software and protocol are also security challenges on the logical layer. The vulnerability of mining and repair, safe programming and security API may be promising solutions. (3) Outside security threats. Attacks to the controller of SDN or NFV may result in the unavailability of service, such as the DDos attack. Security threat detection approaches can solve the issue.

## 3 Hierarchical Attack and Defense Model

The security threat to 5G networks is always caused by a combination of attacks rather than a single attack. For example, to steal the private data of a user, the attacker first breaks the communication between the BS and the user by sending jamming signals, and then establishes communication with the user as a fake BS by exploiting the vulnerability of authorization protocol. Thus, the attacker can obtain all data transmitted through it. By combining signal jamming, vulnerability, fake device, a threat is formed to the user. We call the combination of multiple attacks an attack chain. The attack chain always brings a serious security threat for 5G networks. However, the attack chain of the 5G network receives little attention. In this section, a hierarchical attack and defense model of the 5G network is proposed, such that the attack chain and correspond defense strategies can be obtained based on this model.

As mentioned in the previous section, the 5G architecture is summarized to a hierarchical framework. Each layer of the framework has its own attack/defense objects and corresponding relationships between these objects. Note that, the object in this section is a specific object rather than a rough division in the previous section, e.g., the object OS should be specified with a vendor and detailed version. Based on the hierarchical framework, a graph (Figure 2(a)) can be constructed on each layer with the object as a node and the relationship between any two objects as an edge. For example, two objects on the physical layer can establish an edge on the graph according to the connectivity relationship on the topology and two nodes on the logical layer can establish an edge according to the logical relationship of function between the two objects. The relationship can be viewed as the edge attribute, and each edge may have multiple attributes. Between layers, objects can also be connected by vertical edges according to relationships such as functional support, resource sharing, management, orchestration, and so on. Thus, the graph of each layer is connected by the vertical edge and forms a hierarchical graph, as shown in Figure 2(b). For the convenience of illustration, the sliced service layer is denoted by one slice, so does the corresponding application layer.

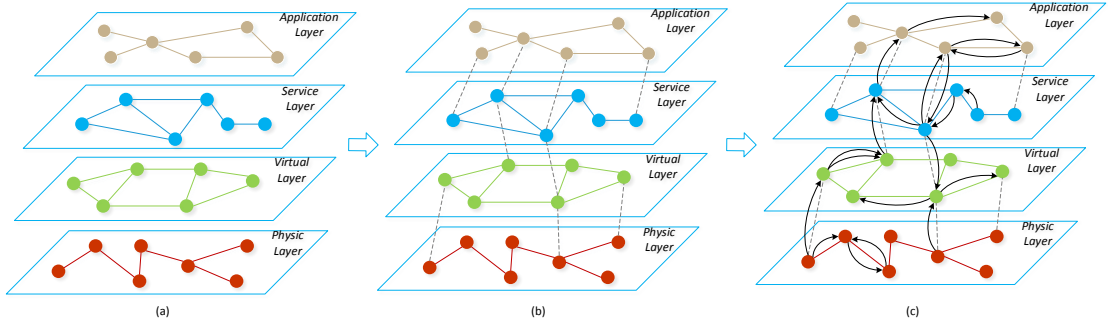

Figure 2. Hierarchical Attack and Defense Model

For an attack to an attack/defense object, it is represented by a triple <*object*, *attack*, *a-results*>. The *object* is the attack/defense object. The *attack* contains two aspects, the attack condition, and the attack method. The *a-results* is a list of attack results and reflects the effect of the attack on different objects. The attack effect can directly correspond to the *object* itself, and it can also correspond to another attack/defense objects connected to the *object* in the hierarchical graph. The *a-results* is represented by a pair <*affected object*, *permission*>, and the *permission* obtained by the attack contains readable, writable, executable, disable, and so on. Based on the triple <*object*, *attack*, *result list*>, a directed edge is established between the attacked object and the affected object to represent the result of an attack. If the attack obtains multiple permissions of the affected object, there may exist multiple directed edges. For example, <$O1$, <$C1$, $A1$>, <<<$O2$, *readable*>, <$O3$, *executable*>>> indicates that the attack $A1$ to the object $O1$ affects two objects $O2$ and $O3$, and obtains readable and executable permissions from the two objects, respectively. Therefore, two directed edges are established from $O1$ to $O2$ and $O3$, respectively.

Based on the object and the attack it suffers, each attacked object generates multiple outward directed edges on the hierarchical graph. All these directed edges can be combined to form a hierarchical directed graph, as shown in Figure 2(c). Based on the directed graph, the attack chain can be obtained. Although some connected edges can form a directed path, the path may not correspond to an effective attack chain. The attack chain is not a simple mapping from the directed path to the chain. Therefore, to find an effective directed path, for an edge and its adjacent edge, we first determine whether the *permission* obtained by the attack on the edge can satisfy the *attack condition* of the attack on the adjacent edge. If the condition is satisfied, the two edges are connected to form an effective directed path, i.e., an attack chain. Otherwise, the directed path is not an attack chain.

The defense strategy is denoted by a triple <*cost*, *method*, *d-results*>, and it is viewed as the attribute of corresponding directed edge. The *cost* is the condition and resource to perform the defense strategy. The *method* is the accurate defense method. The *d-result* is the effect of the defense method. Thus, each attack chain corresponds to a list of defense strategies. There are several defense options to deal with the attack chain: choose a defense strategy for each attack on the attack chain; choose several defense strategies according to specific security requirements or resource constraints; choose a defense strategy for the critical attack on the chain to break the connectivity of attack chain.

A hierarchical attack and defense model of the 5G network is constructed based on the hierarchical directed graph. Security threats in 5G networks are no longer limited to an attack but extend to intra-layer and inter-layer by combining multiple attacks. The hierarchical model brings two advantages. On the one hand, current security threats are analyzed deeply such that effective strategies to resist the security threat can be found. On the other hand, potential security threats can also be found based on directed paths such that the security risk can be eliminated as soon as possible.

## 4 Automated Attack and Defense Framework

Based on the hierarchical attack and defense model, the research on 5G security can be carried out from two aspects: the attack and the defense. However, most of the 5G security researches require domain knowledge and manual interventions, which is difficult to meet the security requirements regarding coverage, accuracy, and timeliness for security threats. Therefore, automated attack and defense becomes the key research to solve 5G security issues. In this section, we propose an automated attack and defense framework for 5G networks.

The automated attack and defense of 5G network consist of four parts: the security-related data, automated attack technologies, automated defense technologies, and the security testbed. The automated attack and defense framework is shown in Figure 3. A 5G security knowledge graph is first constructed based on massive security data of 5G networks. By using the knowledge graph, a hierarchical directed graph is constructed to support the automated attack and defense technologies. Then, automated attack technologies and automated defense technologies are studied to find existing, and potential security threats and similar defense strategies, and the results can feedback to the knowledge graph. In the end, the security testbed is studied to provide a verification platform to new automated attack and defense technologies, attack methods and defense strategies.

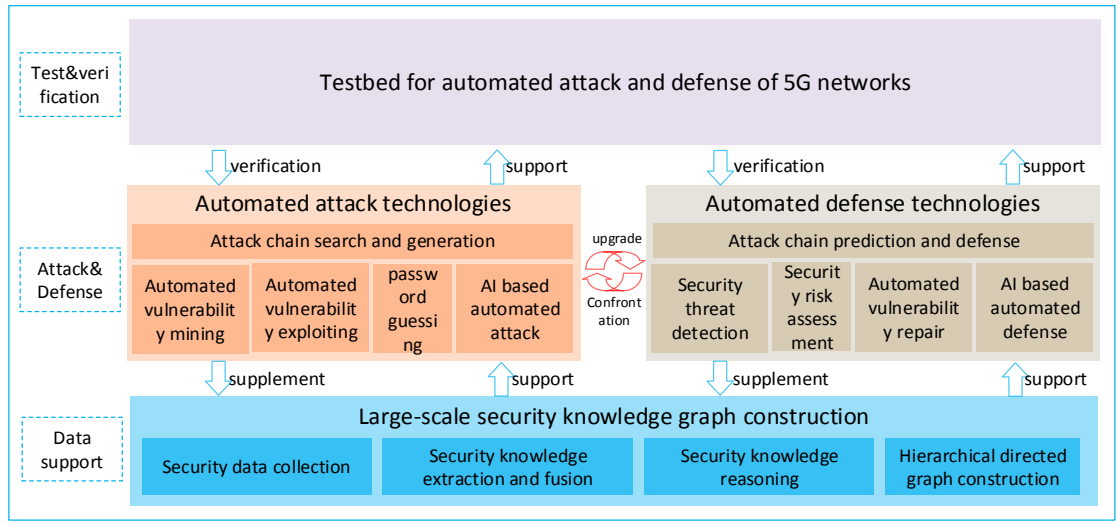

Figure 3. Automated Attack and Defense Framework

**4.1 Security knowledge graph construction**

The security knowledge graph collects scattered security knowledge and integrates the knowledge into a graph. According to the hierarchical attack and defense model, the security knowledge graph should contain multiple entities with corresponding attributes and relationships, such that the information on the graph used to construct the hierarchal directed graph can be satisfied. Figure 4 shows an example of security knowledge graph. Some entities, relationships, and attributes, which do not have a direct relationship with the attack and defense model, are not shown in the figure.

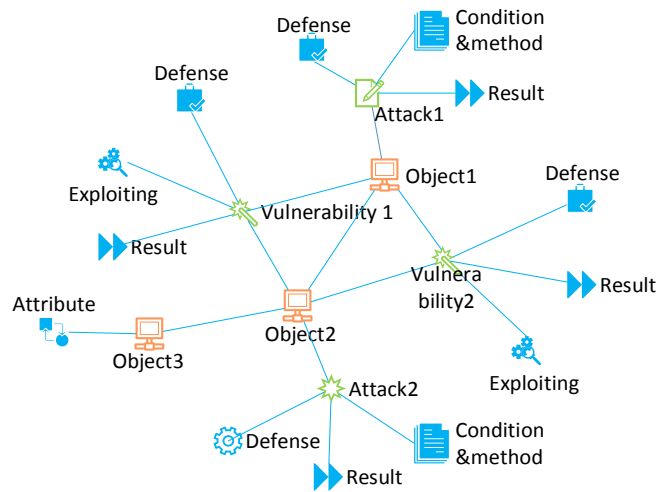

Figure 4. An Example of Security Knowledge Graph

The construction of security knowledge graph is divided into three steps: security data collection, security knowledge extraction, and fusion, security knowledge reasoning. Though there are already some mature technologies that can be used directly for knowledge graph, some special requirements for 5G networks should be satisfied. The security data collection should satisfy large-scale and dynamic requirements. The security data is available in a large number of channels, such as 0day, exploit-db, Github, scap, dark network, Shodan, etc. Meanwhile, the data is updated frequently. Efficient web crawling technology and increment-based effective synchronization mechanism are two promising methods. The main challenges for data extraction and fusion are completeness and accuracy. Since the security data is collected from multiple sources and it is not always structured, the entity and relationship may not be completely extracted and different representations of the same entity may not be correctly identified. Semi-structured and unstructured data extraction and multi-source knowledge fusion should be further studied. Security knowledge reasoning focuses on the efficiency, i.e., how to find the hidden relationship from the knowledge graph efficiently. Since the knowledge graph is so large that it is difficult to satisfy the requirement. Subgraph-based knowledge reasoning or AI-based knowledge reasoning may be promising approaches.

Based on the security knowledge graph, the hierarchical directed graph can be constructed. Nodes on the hierarchical graph are the objects on the knowledge graph, and edges are the entities, such as attack, vulnerability, which connects to the object on the knowledge graph. The layer information corresponds to the attribute of the object on the knowledge graph.

**4.2 Automated Attack technology**

The automated attack is studied in two aspects. (1) Key automated attack technologies are studied separately, including automatic vulnerability mining, automated vulnerability exploiting, password guessing, AI-based automated attack, etc. These technologies are always focused on one type of attack. (2) Attack chain search and generation technology are used to find existing attack chains and potential attack chains by combining the key automated attack technologies on the hierarchical directed graph. The combination of the above technologies forms an integrated automatic attack technology.

The automated vulnerability mining technology finds vulnerabilities from software or protocols automatically. The fuzzing technique is combining with symbolic execution technique, such as AFL and its improved approaches [11], are always used. The under-reporting and efficiency are two challenges for automatic vulnerability mining. Researches on high-quality seed generation, memory recovery, memory error detection may be helpful to overcome these challenges. The automated vulnerability exploiting technology will automatically locates the return address and layout memory to execute shellcode based on the vulnerability information. A Data-oriented exploit using the data flow stitching technique to hijack the program's control flow [12] is a promising technology, but the success rate and universality issues should be solved. For the password guessing technology, AI-based technologies like PassGAN are used to generate high-quality passwords for password library. Based on the password library, password guessing tools, such as HashCat, John the Ripper, are used to guess the password. The quality of the password library and guessing efficiency are two challenges for the password guessing technology. Correspondingly, the deep learning and parallel computing technology on CUP-GPU can be further studied. The AI-based automated attack is always used to bypass security detection and defense by dynamic attack methods, e.g., the DDos attack to the BS can bypass security detection by changing the character and the rule of device access with the AI technology.

The attack chain search technology finds an effective attack chain from existing attack chains on the hierarchical directed graph. Given an object, an effective attack chain starting from this object can be obtained by searching a path according to certain conditions, such as the minimal cost, the maximum threat. An effective attack chain targeting this object can also be obtained according to some conditions. The efficiency is the main focus of attack chain search technology. The path search technology on a big graph can be studied.

The attack chain generation technology constructs an effective attack chain based on the directed path on the hierarchical directed graph. A potential attack chain should be first found to generate a new attack chain, then the corresponding attack method of each edge should be found. The AI technology can be used to find the potential attack chain by learning the characteristics of existing attack chains. For the potential attack chain, above key automated attack technologies and the security knowledge graph can be used to find to the particular attack method for each edge.

**4.3 Automated Defense Technology**

The design of automated defense technology is similar to that of automated attack technology, and it also consists of two parts. (1) Key technologies for automated defense, including security threat detection, security risk assessment, automated vulnerability repair, AI-based automated defense, etc. (2) The attack chain prediction and defense. It is used to predict the attack chain and take corresponding defense strategy with the support of key technologies.

The security threat detection technology is used to detect the attack on objects of 5G networks. The security situational awareness is a promising approach. However, for 5G networks, security threats come from multiple layers, and traditional situational awareness may be limited. Therefore, the situational awareness technology that supports NFV, SDN and network slicing and awares changes in physical layer requires further research. The security risk assessment technology quantitatively evaluates security threat, defense strategy and the cost of defense, such that the choice of the defense strategy is effectively supported. Due to a large number of objects and the complex 5G network environment, the risk assessment of the 5G network should be improved based on previous assessment approaches. The automated vulnerability repair technology provides automated repair strategies for vulnerabilities, such as automated patching, search-based program repair, semantics-based program repair [13]. The AI-based automated defense technology, which corresponds to the AI based attack, learns dynamic defense methods from existing defense or attack.

The attack chain prediction and defense technologies rely to the support of the above key technologies. By detecting the attack to an object, the attack path (i.e., attack chain) and attack objects can be predicted, and corresponding defense strategies are taken. Since the process of attack and defense is dynamic, the game theory can be used for the prediction of attack chain and the choice of defense strategy. Meanwhile, reinforcement learning learns the best reward action through trial and error, which is helpful to the choice of defense strategy. Thus, the seamless integration of game theory and reinforcement learning may be a promising approach for attack chain prediction and defense.

**4.4 Testbed of 5G Security**

The 5G testbed is an important part for the automated attack and defense framework. For one thing, the new security technologies can be verified on the testbed. For another, it is impossible to perform an attack and defense experiment on the real network even if the 5G network has been deployed. Based on the testbed, current security threats can be deeply studied, and potential security threats can be detected and effectively responded.

A variety of testbeds for 5G networks are currently proposed, but the testbed for 5G security is still lacking, especially for the automatic attack and defense research. The 5G security testbed can be studied in two aspects. (1) A unified testbed framework for large-scale, flexible and realistic security testing can be built. Based on the unified testbed framework, the simulation of each layer can be gradually implemented. (2) A special security testbed of a layer, such as the physical layer security testbed, the virtual layer security testbed, can be studied for the security research on the corresponding layer. The testbed should also take into account the security relationship with other layers. (3) The testbed for vertical services can also be studied. Since the 5G network provides a variety of services and applications, the testbed can simulate and reproduce the security scenario for a specific service, e.g., the smart city, smart campus [14].

## 5 Conclusion

The 5G security has become one of the important research fields for the 5G network. In this article, the physical layer security and the logical layer security are integrated, and 5G security is studied in the perspective of automated attack and defense. We first propose a universal hierarchical 5G framework and discuss the security challenge of the 5G object. Based on the framework, a hierarchical attack and defense model is detailed to describe the attack and defense of intra-layer and inter-layer. Based on the model, an automated attack, and defense framework is presented with four key technologies, including knowledge graph construction, automated attack, automated defense and security testbed, to provide accurate and fast security strategies for security threats of 5G network.

## Acknowledgements

This work is funded by the National Natural Science Foundation of China (No. 61871140, 61702223, 61702220, 61572153, U1636215) and the National Key Research and Development Program of China (No. 2018YFB0803504).

**Zhihong Tian** Ph.D., professor, PHD supervisor. Dean of cyberspace institute of advanced technology, Guangzhou University. Standing director of CyberSecurity Association of China. Member of China Computer Federation. From 2003 to 2016, he worked at Harbin Institute of Technology. His current research interest is computer network and network security.

**Yanbin Sun** received the B.S., M.S. and Ph.D degree in Computer Science from Harbin Institute of Technology (HIT), Harbin, China. He is currently an assistant professor in Guangzhou University, China. His research interests include network security, future networking and scalable routing.

**Shen Su** born in 1985, Ph.D., assistant professor, Guangzhou Unversity. His current research interest is inter-domain routing and security.

**Mohan Li** received her B.S., M.S. and Ph.D degree in Computer Science from Harbin Institute of Technology (HIT), Harbin, China. She is currently an assistant professor in Guangzhou University, China. Her research interests include data quality and data security.

**Xiaojiang(James) Du** (M'03–SM'09) received the B.E. degree from Tsinghua University, China, in 1996, and the M.S. and Ph.D. degrees from the University of Maryland, College Park, in 2002 and 2003, respectively, all in electrical engineering. He is currently a Professor with the Department of Computer and Information Sciences, Temple University. His research interests are security, systems, wireless networks, and computer networks. He has published over 240 journal and conference papers in these areas. He serves on the editorial boards of two international journals.

**Mohsen Guizani** (S'85–M'89–SM'99–F'09) received the bachelor's (Hons.) and master's degrees in electrical engineering and the master's and Ph.D. degrees in computer engineering from Syracuse University, Syracuse, NY, USA, in 1984, 1986, 1987, and 1990, respectively. He is currently a Professor and the ECE Department Chair with the University of Idaho. His research interests include wireless communications and mobile computing, computer networks, mobile cloud computing, security, and smart grid. He was the Chair of the IEEE Communications. Society Wireless Technical Committee and the TAOS Technical Committee. He currently serves on the editorial boards of several international technical journals. He is the Founder and the Editor-in-Chief of Wireless Communications and Mobile Computing (Wiley). He served as the IEEE Computer Society Distinguished Speaker from 2003 to 2005.